\documentclass[11pt,letterpaper]{article}
\pdfoutput=1
\usepackage{jheppub}
\usepackage{amsfonts, amsthm,mathrsfs}
\usepackage[english]{babel}
\usepackage[utf8]{inputenc}
\usepackage{slashed}
\hypersetup{unicode}

\newcommand{\eq}{\begin{equation}}
\newcommand{\feq}{\end{equation}}
\newcommand{\eqn}{\begin{eqnarray}}
\newcommand{\feqn}{\end{eqnarray}}

\newcommand{\ma}[1]{\mbox{$\mathcal{#1}$}}
\newcommand{\mas}[1]{\mbox{$\mathscr{#1}$}}
\newcommand{\D}{{\rm d}}

\newcommand{\ti}{\tilde}

\title{Hairy black holes in $N=2$ gauged supergravity}

\author{Federico Faedo$^a$,}
\author{Dietmar Klemm$^{a,b}$}
\author{and Masato Nozawa$^{a,b}$}
\affiliation{$^a$ Dipartimento di Fisica, Universit\`a di Milano, \\
\hspace*{0.15cm} Via Celoria 16, 20133 Milano, Italy. \\
$^b$ INFN, Sezione di Milano, \\
\hspace*{0.15cm} Via Celoria 16, 20133 Milano, Italy.
}
\emailAdd{federicomichele.faedo@studenti.unimi.it}
\emailAdd{dietmar.klemm@mi.infn.it}
\emailAdd{masato.nozawa@mi.infn.it}
\preprint{IFUM-1040-FT}

\abstract{We construct black holes with scalar hair in a wide class of four-dimensional $N=2$
Fayet-Iliopoulos gauged supergravity theories that are characterized by a prepotential containing
one free parameter. Considering the truncated model in which only a single real scalar survives,
the theory is reduced to an Einstein-scalar system with a potential, which admits at most two AdS critical points and is expressed in terms of a real superpotential. Our solution is static, admits maximally symmetric horizons, asymptotically tends to AdS space corresponding to an extremum of the superpotential, but is disconnected from the Schwarzschild-AdS family. The condition under which the spacetime admits an event horizon is addressed for each horizon topology. It turns out that for hyperbolic horizons the black holes
can be extremal. In this case, the near-horizon geometry is $\text{AdS}_2\times\text{H}^2$, where
the scalar goes to the other, non-supersymmetric, critical point of the potential.
Our solution displays fall-off behaviours different from the standard one, due to the fact that the mass parameter $m^2=-2\ell^{-2}$ at the supersymmetric vacuum lies in a characteristic range $m^2_{\rm BF}\le m^2\le m^2_{\rm BF}+\ell^{-2}$ for which the slowly decaying scalar field is also normalizable ($m^2_{\rm BF}=-9/(4\ell^2)$ denotes the Breitenlohner-Freedman bound). Nevertheless, we identify a well-defined mass for our spacetime, following the prescription of Hertog and Maeda. Quite remarkably, the product
of all horizon areas is not given in terms of the asymptotic cosmological constant alone, as one would
expect in absence of electromagnetic charges and angular momentum.
Our solution shows qualitatively the same thermodynamic behaviour as the Schwarzschild-AdS black hole, but the entropy is always smaller for a given mass and AdS curvature radius. We also find that our spherical black holes are unstable against radial perturbations. 
}


\begin{document}
\maketitle
\flushbottom

\section{Introduction}

The uniqueness theorems for stationary, nonextremal black holes in the Einstein-Maxwell
system \cite{Israel:1967wq,Israel:1967za,Carter:1971zc,Robinson:1975bv,Mazur:1982db} are one of
the crowning triumphs of general relativity. Stationary black holes in asymptotically flat spacetimes are
thus completely specified by the asymptotic charges ($M, J, Q$) and exhausted by the Kerr-Newman family. One might expect the validity of this statement as stemming from the observation that the higher multipole moments present at the formation of black holes would die away due to electromagnetic and gravitational radiation. A perturbative analysis of black-hole ringdowns affirmatively supports this
belief \cite{Price:1971fb}. Inspired by these works, Ruffini and Wheeler proposed a novel
conjecture \cite{Ruffini:1971bza} that black holes in more general settings do not allow additional `hair' to be characterized. 

Unlike the Einstein-Maxwell system without a cosmological constant, the global boundary value problem utilizing a nonlinear sigma model cannot be adopted in an Einstein-scalar system if the scalar fields have a potential. This difficulty restricts the applicability of the no scalar-hair proof only to the static case. When the potential $V(\phi)$ of a scalar field satisfies $\phi\partial V/\partial \phi\ge 0$, Bekenstein gave an elegant proof which rules out nontrivial scalar configurations outside an  asymptotically flat  static black hole with a regular horizon \cite{Bekenstein:1971hc,Bekenstein:1972ny}. This theorem was later generalized to an arbitrary nonnegative potential \cite{Heusler:1992ss,Sudarsky:1995zg} and to noncanonical scalar
systems \cite{Graham:2014mda}. By sidestepping some assumptions that go into these theorems,
black holes are not necessarily getting bald. Prototype examples are black holes sourced by a
conformally coupled scalar
field \cite{Bekenstein:1974sf,Bekenstein:1975ts} and the `coloured' black holes dressed with a Yang-Mills
field \cite{Bizon:1990sr}. Unfortunately, both of these solutions are
unstable \cite{Bronnikov:1978mx,Bizon:1991nt} and are not realizable as final states of the gravitational collapse. Note that, based on the results of \cite{Huebscher:2007hj,Hubscher:2008yz},
refs.~\cite{Meessen:2008kb,Bueno:2014mea,Meessen:2015nla} analytically constructed coloured hairy
BPS black holes. Being supersymmetric, these solutions are expected to be stable\footnote{However, 
since they admit degenerate horizons, these black holes generically suffer from another kind of instability found in \cite{Aretakis:2011ha,Lucietti:2012sf}.}. Black holes in four-dimensional $N=2$ ungauged
supergravity violating the no-hair conjecture were found in \cite{Bueno:2013vua}, but unfortunately
the special Kähler metric of the scalar field target space is not positive definite when evaluated on the
solutions, and thus ghost modes appear.

In asymptotically AdS spacetimes, the situation changes drastically and the story is much richer. Although
a no-hair theorem for static and spherically symmetric asymptotically AdS black holes has been
established for nonconvex potentials \cite{Sudarsky:2002mk},  one cannot validate the uniqueness of
the Schwarzschild-AdS black hole without additional restrictions even in the Einstein-$\Lambda(<0)$ system~\cite{Anderson:2002xb}. Moreover, finding solutions itself is a formidable task, since the usual
solution-generating techniques do not work in the presence of scalar potentials. Hence, we do not yet
grasp the whole picture of the solution space of AdS black holes. Recently, many black holes admitting
scalar hair have been obtained by ansatz-based approaches, for which the potential is `derived' in such a
way that the assumed metric solves the equations of
motion \cite{Anabalon:2012sn,Anabalon:2012ta,Anabalon:2013qua,Gonzalez:2013aca,
Gonzalez:2014tga,Cadoni:2015gfa}. This kind of heuristic approach gives in general a peculiar form of the
scalar potential, which lacks physical motivations unless the parameters are tuned appropriately. 

Here we are interested in black holes with scalar hair in gauged supergravities. In this framework, the scalar
fields acquire a potential due to the gaugings, which leads us naturally to set up the situation of
asymptotically AdS spacetimes. AdS black holes with scalar hair are of primary importance in the context
of the gauge/gravity correspondence and applications to condensed matter physics. From the holographic
point of view, the excitation of unstable modes creates a bound state of a boundary
tachyon \cite{Gubser:2000mm}. It follows that the instability of a hairy black hole provides an interesting
phase of the dual field theories. 

In this paper, we construct a static, neutral black hole with a maximally symmetric horizon admitting
scalar hair in $N=2$ supergravity with Fayet-Iliopoulos gauging. We consider a model with a single vector
multiplet for which the prepotential involves a single free parameter. By truncating to the subsector of a
real scalar field and vanishing gauge fields, the resulting scalar potential can be expressed in terms of a
superpotential. One of the critical points extremizes also the  superpotential and the mass parameter at
the critical point is given by $m^2=-2\ell^{-2}$, where $\ell $ is the AdS curvature radius. It is worth
noting that the mass is in the characteristic range $m^2_{\rm BF}\le m^2\le m_{\rm BF}^2+\ell^{-2}$,
where $m^2_{\rm BF}=-9/(4\ell^2)$ denotes the BF bound \cite{Breitenlohner:1982jf} under which the
scalar field is perturbatively unstable.  In this case, the slowly decaying mode of the scalar field is also
normalizable, for which the asymptotic fall-off behavior deviates from the standard one. It then follows that
the conventional methods for computing conserved quantities in asymptotically AdS
spacetimes \cite{Abbott:1981ff,Ashtekar:1984zz,Henneaux:1985tv,Hollands:2005wt,Katz:1996nr} cannot
be applied. In spite of this, some authors have used without justification a formula which is valid only in the
case of Dirichlet boundary conditions. We exploit the prescription of Hertog and Maeda \cite{Hertog:2004dr}
to compute conserved charges valid also for the mixed boundary conditions. We explore in detail the
condition under which our solution admits an event horizon for each horizon topology.  We also analyze
the Wick rotation of our solution, which describes an asymptotically de Sitter black hole.

An outline of the present paper consists as follows. In section \ref{sec:lagrangian}, we give a brief review
of $N=2$ gauged supergravity with Abelian Fayet-Iliopoulos gauging. We truncate the model down to a
single scalar and examine the structure of the scalar potential. In section~\ref{sec:hairysolution}, we present
the hairy black hole solution and show that some hairy black holes obtained in the literature are recovered
by taking certain limits. In section \ref{sec:physicaldiscussion}, we address some physical properties of our
solution. We identify a well-defined mass function of the spacetime, explore the structure of the Killing
horizons in detail, and work out the conditions under which the solution admits an event horizon. We then
investigate the thermodynamic behaviour of the black holes and discuss their (in)stability. An extension to
the asymptotically de Sitter case is also given. Section \ref{final-rem} concludes with some remarks.

\section{Fayet-Iliopoulos gauged $N=2$, $D=4$ supergravity}
\label{sec:lagrangian}

We consider $N=2$, $D=4$ gauged supergravity coupled to $n_V$ abelian
vector multiplets \cite{Andrianopoli:1996cm}\footnote{Throughout this paper,
we use the notations and conventions of \cite{Vambroes}.}.
In addition to the vierbein $e^a{}_{\mu}$, the bosonic field content consists of the
vectors $A^I_{\mu}$ enumerated by $I=0,\ldots,n_V$, and the complex scalars $z^{\alpha}$
($\alpha=1,\ldots,n_V$). These scalars parametrize
a special K\"ahler manifold, i.e., an $n_V$-dimensional
Hodge-K\"ahler manifold which is the base of a symplectic bundle characterized by the
covariantly holomorphic sections
\begin{equation}
{\cal V} = \left(\begin{array}{c} X^I \\ F_I\end{array}\right)\,, \qquad
{\cal D}_{\bar\alpha}{\cal V} = \partial_{\bar\alpha}{\cal V}-\frac 12
(\partial_{\bar\alpha}{\cal K}){\cal V}=0\,, \label{sympl-vec}
\end{equation}
where ${\cal K}={\cal K}(z^\alpha, \bar z^\alpha) $ is the K\"ahler potential and ${\cal D}_\alpha$ denotes the K\"ahler-covariant ${\rm U}(1)$ derivative. 
The covariantly symplectic section ${\cal V}$ obeys the symplectic constraint
\begin{equation}
\langle {\cal V}\,,\bar{\cal V}\rangle = X^I\bar F_I-F_I\bar X^I=i\,,
\end{equation}
where $\langle\,,\, \rangle$ denotes the symplectic inner product. 
It is also useful to define $v(z)$ as
\begin{equation}
{\cal V}=e^{{\cal K}(z,\bar z)/2}v(z)\,,
\end{equation}
where $v(z)$ corresponds to a holomorphic symplectic vector,
\begin{equation}
v(z) = \left(\begin{array}{c} Z^I(z) \\ \frac{\partial}{\partial Z^I}F(Z)
\end{array}\right)\,.
\end{equation}
$F$ is a homogeneous function of degree two, referred to as the prepotential,
whose existence is assumed to get the final expression.
In terms of $v$,  one finds the K\"ahler potential
\begin{equation}
e^{-{\cal K}(z,\bar z)} = -i\langle v\,,\bar v\rangle\,.
\end{equation}
The matrix ${\cal N}_{IJ}$ describes the coupling between the scalars
$z^{\alpha}$ and the vectors $A^I_{\mu}$, and is defined by the relations
\begin{equation}\label{defN}
F_I = {\cal N}_{IJ}X^J\,, \qquad {\cal D}_{\bar\alpha}\bar F_I = {\cal N}_{IJ}
{\cal D}_{\bar\alpha}\bar X^J\,.
\end{equation}
The bosonic Lagrangian reads
\begin{eqnarray}
e^{-1}{\cal L}_{\text{bos}} &=& \frac 12R + \frac 14(\text{Im}\,
{\cal N})_{IJ}F^I_{\mu\nu}F^{J\mu\nu} - \frac 18(\text{Re}\,{\cal N})_{IJ}\,e^{-1}
\epsilon^{\mu\nu\rho\sigma}F^I_{\mu\nu}F^J_{\rho\sigma} \nonumber \\
&& -g_{\alpha\bar\beta}\partial_{\mu}z^{\alpha}\partial^{\mu}\bar z^{\bar\beta}
- V\,, \label{Lagrangian_gen}
\end{eqnarray}
with the scalar potential
\eq
V = -2g^2\xi_I\xi_J[(\text{Im}\,{\cal N})^{-1|IJ}+8\bar X^IX^J]\,, \label{pot_gen}
\feq
that results from U$(1)$ Fayet-Iliopoulos gauging. Here, $g$ denotes the
gauge coupling and the $\xi_I$ are constants. In what follows, we define
$g_I=g\xi_I$.

In this paper, we focus on a model with prepotential characterized by a single parameter $n$,
\begin{align}
\label{}
F(X) = - \frac{i}{4} ({X^0})^{n} ({X^1})^{2-n}\,, 
\end{align}
that has $n_V=1$ (one vector multiplet), and thus just one complex scalar $z$.
This is a truncation of the stu model with $F\propto(X^0X^1X^2X^3)^{1/2}$
(set $X^2={X^0}^{2n-1}$, $X^3={X^1}^{3-2n}$). Note that, for zero axions and a special choice of
the FI parameters $\xi_I$, the latter can be obtained by dimensional reduction from eleven-dimensional
supergravity~\cite{Cvetic:1999xp}.

Choosing $Z^0=1$, $Z^1=z$, the symplectic vector $v$ becomes
\begin{equation}
v = \left(\begin{array}{c}
1			\\
z			\\
-\frac{i}{4} \, n z^{2-n}	\\
-\frac{i}{4} \, (2-n) z^{1-n}
\end{array}\right)\,.
\end{equation}
The K\"ahler potential is given by
\begin{equation}
e^{-\mathcal{K}} = \frac{1}{4} \left[n(z^{2-n} + \bar{z}^{2-n}) + (2-n)(z^{1-n}\bar{z} + z\bar{z}^{1-n})
\right]\,.
\end{equation}
When $n=1$, the scalar manifold describes ${\rm SU}(1,1)/{\rm U}(1)$.  
In what follows, we shall restrict to the truncated model with a single real scalar $z=\bar z$. 
In that case, the metric and kinetic matrix for the vectors boil down to
\begin{displaymath}
{g_{z\bar{z}}|}_{\text{Im}z=0} = {\partial_z \partial_{\bar{z}} \mathcal{K}|}_{\text{Im}z=0} =
\frac{n(2-n)}{4 z^2}\,, \qquad {{\cal N}|}_{\text{Im}z=0} = -\frac i4\left(\begin{array}{cc}
n z^{2-n} & 0 \\ 0 & (2-n)z^{-n}\end{array}\right)\,,
\end{displaymath}
while the potential \eqref{pot_gen} becomes
\begin{equation}
V = - 8\left[ \frac{2n-1}{n} \, g_0^2 \, z^{n-2} + 4 \, g_0 g_1 \, z^{n-1} + \frac{3-2n}{2-n} \, g_1^2 \, z^n \right] \ .
\end{equation}
Moreover, since we are interested in uncharged black holes, we set $F^I_{\mu\nu}=0$. Then the action
reduces to
\begin{equation}
\label{action}
S = \int \left[\frac R2 - \frac12 g^{\mu\nu} \partial_\mu\phi\,\partial_\nu\phi - V(\phi) \right]\sqrt{-g}\,
\D^4 x\,,
\end{equation}
where we defined the canonical scalar field $\phi$ by $\phi = \lambda_n^{-1} \ln z$, with
\begin{equation}
\lambda_n = \sqrt{\frac{2}{n(2-n)}}\,, \qquad 0<n<2\,.
\end{equation}
Here the allowed range of $n$ comes from the restriction ${\rm Im}\,\ma N<0$, 
which assures positivity of the kinetic term for the gauge fields. In terms of $\phi$,
the potential is given by
\begin{equation}
\label{V}
V(\phi) = - 8\left[\frac{2n-1}{n} \, g_0^2 \, e^{\lambda_n (n-2) \, \phi} + 4 \, g_0 g_1 \, e^{\lambda_n (n-1) \, \phi} + \frac{3-2n}{2-n} \, g_1^2 \, e^{\lambda_n n \, \phi} \right]\,,
\end{equation}
which can be written in terms of a superpotential $W=W(\phi)$ as
\begin{equation}
\label{superpotential}
V = 4\left[2(\partial_\phi W)^2 - 3W^2\right]\,,
\end{equation}
where
\begin{equation}
\label{W}
W(\phi) = g_1 e^{n\lambda_n\phi/2} + g_0 e^{(n-2)\lambda_n\phi/2}\,.
\end{equation}
In what follows we shall assume that both $g_0$ and $g_1$ are positive. 
Remark that the theory is invariant under
\begin{align}
\label{}
g_0 \to g_1\,, \qquad g_1 \to g_0 \,, \qquad n \to 2-n \,, \qquad \phi \to -\phi \,. 
\end{align}
This invariance allows us to restrict to the range $0<n\le 1$ for the discussion of the physical properties
of the solution. In spite of this, we shall consider the full range $0<n<2$ for clarity of our argument. 

\begin{figure}[t]
\begin{center}
\includegraphics[width=15cm]{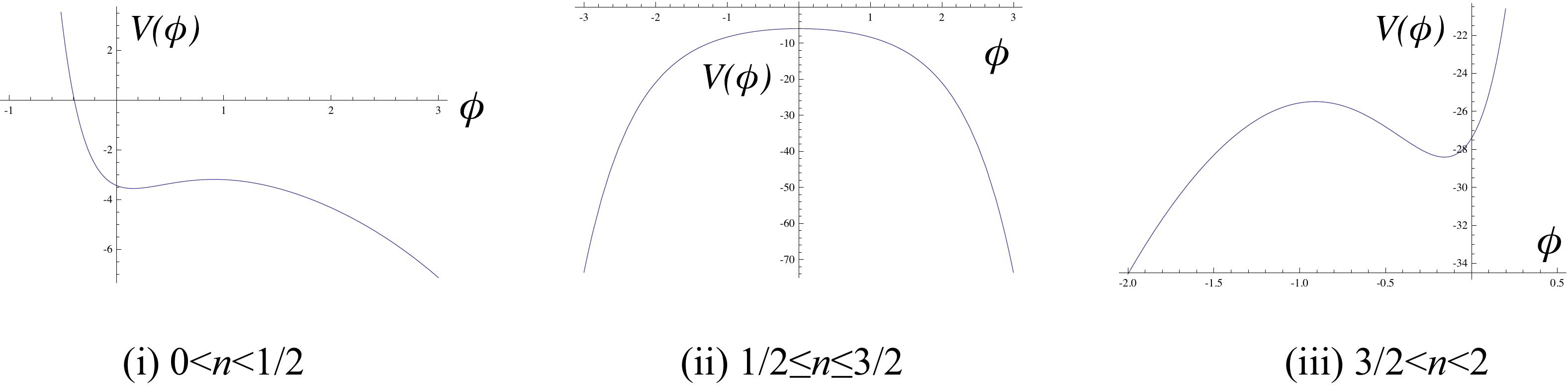}
\end{center}
\caption{The behaviour of the scalar potential for (i) $0<n<1/2$, (ii) $1/2<n<3/2$ and (iii) $3/2<n<2$. 
The $1/2\le n<3/2$ case admits only a global maximum corresponding to the supersymmetric vacuum, otherwise there exist two critical points. The extrema always correspond to a negative cosmological
constant.}
\label{fig:V}
\end{figure}

One finds that the potential \eqref{V} has two critical points (see fig.~\ref{fig:V}), namely
\begin{align}
\label{extrema}
e^{\lambda_n \phi_1}=\frac{g_0(2-n)}{g_1 n}\,, \qquad 
e^{\lambda_n \phi_2} =\frac{g_0(2-n)(1-2n)}{g_1(3-2n)n} \,. 
\end{align} 
$\phi=\phi_1$ always exists for $0<n<2$. 
Since the extremum $\phi=\phi_1$ is also a critical point of the superpotential, 
it describes a supersymmetric vacuum. On the other hand, 
the extremum $\phi=\phi_2$ does not exist in the range $1/2\le n\le 3/2$ 
and it breaks supersymmetry.

If we define $\phi\equiv\phi_1+\delta\phi$ and expand the potential around $\phi_1$, the action
\eqref{action} can be written as
\begin{equation}
S = \int \left[\frac{R - 2 \Lambda}2 - \frac12 g^{\mu\nu}\partial_\mu\delta\phi\,\partial_\nu\delta\phi -
\frac12 m^2\delta\phi^2 + {\cal O}(\delta\phi^3)\right]\sqrt{-g}\, \D^4 x\,,
\end{equation}
with the cosmological constant $\Lambda=-3\ell^{-2}$, where the asymptotic AdS
curvature radius is given by
\begin{align}
\label{rho0}
\ell=\frac{\rho_0}{2\sqrt 2 g_0} \,, \qquad 
\rho_0 \equiv \frac n{\sqrt2}\left(\frac{(2-n)g_0}{ng_1}
\right)^{1-n/2}\,.
\end{align}
The dimensionless parameter $\rho_0$ was introduced for later convenience. 
The mass parameter $m$ measures the curvature of the potential at the supersymmetric critical point 
and reads
\begin{align}
\label{mass}
  m^2=-2\ell^{-2} \,.
\end{align}
This is exactly the value for a conformally coupled scalar field in AdS.  
One can follow the same steps to show that the other vacuum $\phi=\phi_2$ 
also corresponds to the negative cosmological constant $\Lambda=-3\ell_2^2$ and its mass spectrum 
is given by 
\begin{align}
\label{}
m^2=\frac{6}{\ell_2^2} \,, \qquad 
\ell_2=\frac{n(3-2n)}{4g_0(1-n)}
  \left(\frac{g_0(2-n)(1-2n)}{g_1n(3-2n)}\right)^{1-n/2} \,. 
\end{align}
The supersymmetric vacua ($\partial_\phi W=0$) are always stable since the mass is above or equal to 
the Breitenlohner-Freedman (BF) bound~\cite{Breitenlohner:1982jf}, which can be grasped as
\begin{align}
\label{}
m^2=\partial_\phi^2 V=8 \left[2\left(\partial_\phi^2 W -\frac 34 W\right)^2-\frac{9}8 W^2\right]
\ge -9W^2 =m_{\rm BF}^2 \,,  
\end{align}
where $m_{\rm BF}^2 =-9/(4\ell^2) $ denotes the BF mass bound and the above mass
(\ref{mass}) indeed exceeds this lower bound. 
Here, it is important to remark that the mass (\ref{mass}) at the vacuum $\phi_1$ lies in the BF
range~\cite{Breitenlohner:1982jf}
\begin{equation}
m^2_{\text{BF}} \le m^2 \le m^2_{\text{BF}} + \frac1{\ell^2}\,. 
\label{BFrange}
\end{equation}
By requiring the conservation and the positivity of a suitable energy functional, Breitenlohner and
Freedman \cite{Breitenlohner:1982jf} found that the slowly decaying mode of the scalar field is also
normalizable when the mass parameter lies in this range. It was later shown by Ishibashi and
Wald \cite{Ishibashi:2004wx} that this is also a necessary condition for stability of the scalar field. They proved that the Hamiltonian operator of various fields admits a positive, self-adjoint  extension if and only if the mass is above or equal to the BF bound (see proposition 3.1 in ref.~\cite{Ishibashi:2004wx}). They also showed that the extension of the Hamiltonian is specified by a single parameter taking values in $\mathbb{R}\text{P}^1\cong\text{S}^1$ when the mass of the scalar field lies in the BF range, and that the possible choices of self-adjoint extension correspond to the freedom of choosing possible boundary conditions specified by that parameter. This parameter appears in the asymptotic behaviour of the scalar field as follows. Let $r$ denote the standard radial coordinate of AdS. Then the free scalar field $\phi$ propagating in AdS behaves at infinity as\footnote{When the BF bound is saturated, the two solutions are
degenerate and there appears a second solution with a logarithmic branch. Since this case does not
appear in our model, we shall not pursue it in this paper. We refer the reader to
e.g.~\cite{Henneaux:2004zi,Amsel:2011km} and references therein for a recent discussion of that case.}
\begin{equation}
\label{scalar_behavior}
\phi \sim \frac{\phi_{\pm}(x^i)}{r^{\lambda_{\pm}}}\,, \qquad \lambda_{\pm} = \frac12(3 \pm
\sqrt{9 + 4m^2\ell^2})\,,
\end{equation}
where $x^i$ are coordinates on the conformal boundary. 
For the vacuum $\phi=\phi_1$, this gives $\lambda_+=2$ and $\lambda_-=1$. 
When $m^2 \ge m^2_{\rm BF}+1/\ell^2$ as in the vacuum $\phi_2$, the allowed boundary condition
is only of Dirichlet type, and thus $\phi_-(x^i)=0$. For the BF range (\ref{BFrange}), 
$\phi_-(x^i)$ can also be nonvanishing and the ratio corresponding to the mixed boundary condition is characterized by a single dimensionless parameter $\alpha$ as~\cite{Hertog:2004dr} 
 \begin{equation}
\label{scalar_behavior2}
\phi \sim \frac{\phi_{-}(x^i)}{r^{\lambda_-}}+\frac{\alpha \phi_-^{\lambda_+/\lambda_-}(x^i)}{r^{\lambda_+}}\,.
\end{equation}
This relation is crucial for determining the mass of the spacetime in a later section.

\section{Black holes with scalar hair}
\label{sec:hairysolution}

In this section, we construct black hole solutions with a nontrivial scalar profile in the Einstein-scalar theory described by the action \eqref{action}. The equations of motion following from \eqref{action} read
\begin{equation}
\label{eom}
\Box\phi = V'(\phi)\,, \qquad R_{\mu\nu} = \partial_\mu\phi\, \partial_\nu\phi + g_{\mu\nu} V(\phi)\,,
\end{equation}
where the potential is given by (\ref{V}). As we are looking for static black holes, we use the ansatz
\begin{equation}
\D s^2 = - \, e^{2X(r)}\D t^2 + e^{-2X(r)}\D r^2 + e^{2Y(r)}\D\Sigma_k^2\,, \qquad \phi = \phi(r)\,,
\end{equation}
where $\D\Sigma_k^2$ is the line element on a two-dimensional space with constant curvature $k$, 
\begin{equation}
\D\Sigma_k^2 = \frac{\D\chi^2}{1 - k\chi^2} + \chi^2\D\varphi^2\,, \qquad k=0,\pm 1\,.
\end{equation}
Introducing the new function $\psi = X + Y$, the equations of motion \eqref{eom} boil down to
\begin{equation}
\label{EOMs}
\begin{split}
& X'' + 2 X' \psi' = - e^{-2X} V(\phi)\,, \\
& \psi'' + 2 \psi'^2 - k e^{-2\psi} = -2 e^{-2X} V(\phi)\,, \\
& \phi'' + 2 \phi' \psi' = e^{-2X} V'(\phi)\,, \\
& X'^2 - \psi'^2 + \frac12 \phi'^2 + k e^{-2\psi} = e^{-2X} V(\phi)\,.
\end{split}
\end{equation}
It is worth noting that the equations \eqref{EOMs} may also be derived from the one-dimensional action
\begin{equation}
\label{1Daction}
S = \int\left[\frac12 e^{2\psi}\left(-X'^2 + \psi'^2 - \frac12 \phi'^2\right) - \frac12 e^{-2X + 2\psi}
V(\phi)\right]\D r\,.
\end{equation}
Since $r$ does not appear explicitely in the Lagrangian, the Hamiltonian $H$ is constant, and from
the last equation of \eqref{EOMs} we see that $H$ coincides with $k/2$.

Inspired by \cite{Klemm:2012yg} we set
\begin{equation}
e^{2Y(r)} = f_1(r)^n f_2(r)^{2-n}\,.
\end{equation}
Then it turns out that a class of solutions to the equations of motion \eqref{EOMs} is given by
\begin{align}
\label{solffexp}
& f_1 = \frac{n}{\sqrt{2}}\left(r + \frac{2\beta}{n}\right)\,, \qquad
f_2 = \frac{2-n}{\sqrt{2}}\frac{g_0}{g_1}\left(r - \frac{2\beta}{2-n}\right)\,, \nonumber \\
& e^{2\psi} = 8g_0^2\left(r + \frac{2\beta}{n}\right) \left(r - \frac{2\beta}{2-n}\right)	
\left(r^2 - \frac{4(1-n)}{n(2-n)}\beta r + 4\frac{5 n^2 - 10 n + 4}{n^2 (2-n)^2}\beta^2 +
\frac{k}{8g_0^2}\right)\,, \nonumber \\
& e^{\lambda_n\phi} = \frac{g_0(2-n)}{g_1n}\,\frac{r - 2\beta/(2-n)}{r + 2\beta/n}\,,
\end{align}
where $\beta$ denotes an arbitrary constant. The metric becomes then
\begin{equation}
\label{DKmetric}
\D s^2 = - \frac{e^{2\psi}}{f_1^n f_2^{2-n}}\D t^2 + \frac{f_1^n f_2^{2-n}}{e^{2\psi}}\D r^2 +
f_1^n f_2^{2-n}\D\Sigma_k^2\,.
\end{equation}
Note that the solution is given in terms of a quartic polynomial $e^{2\psi}$ and two linear functions
$f_1,f_2$, whose powers reflect the expression for the prepotential. This generic structure was first
observed in \cite{Cacciatori:2009iz}.
When $\beta =0$, we recover AdS in static coordinates with a two-dimensional constant curvature
space $\D \Sigma_k^2$. Hence $\beta $ measures the deviation from AdS and is proportional to the
mass of the black hole, as we will see later. 
If we take the limit $g_0 \to 0$ with $g_1/g_0$ kept finite, the potential vanishes and the spacetime reduces to the asymptotically flat metric found in \cite{Janis:1968zz,Wyman:1981bd}, which describes a naked singularity. It follows that our solution does not include asymptotically flat black holes with scalar hair.
It is also worthwhile to  remark that the only way to kill the scalar field is $\beta=0$, hence the solution (\ref{DKmetric}) is disconnected from the (topological) Schwarzschild-AdS family. As we will see, the metric (\ref{DKmetric}) admits a parameter range that allows a regular event horizon with a nontrivial scalar field. (\ref{DKmetric}) provides thus a novel example describing a hairy black hole.

\subsection{Comparison with other literature}

Before going into the details of our solution, it is useful to compare it with other asymptotically AdS
hairy black holes previously constructed in the literature. Some of them turn out to correspond
actually to a subclass of \eqref{solffexp}, \eqref{DKmetric}.

The first class is the special case $n =1/2$ and $k = 1$ presented in \cite{Lu:2013ura}, which
represents an exact solution of the theory obtained by the single scalar truncation of 
${\rm U}(1)^4\subset {\rm SO}(8)$ gauged supergravity. 
Setting $\beta_2 = 0$ and subsequently taking the limit $\beta_1 \to \infty$ in eq.~(2.1) of \cite{Lu:2013ura}, we obtain the metric
\begin{equation}
\label{metric20}
\D s^2 = - (1 + g^2 r^2) \sqrt{f_0}  \D t^2 + \frac{\D  r^2}{(1 + g^2 r^2) \sqrt{f_0}} +  r^2 \sqrt{f_0} \, \D \Omega_2^2 \,, 
\end{equation}
and the scalar field
\begin{equation}
\phi = - \frac{\sqrt{3}}{2} \log f_0 \,,
\end{equation}
where $f_0 = 1 - 2\mu/r$. 
From the expression of the potential we learn that $g^2 =g_0^2\rho_0^{-2}$ and it is straightforward to check that the metric \eqref{metric20} can be put in the form \eqref{DKmetric} through the
transformations $r \to \rho_0\left(r - \frac{4}{3}\beta\right)$, $t \to \rho_0^{-1}t$ and requiring that
$\mu = -\frac{8}{3} \beta\rho_0$. 

If one takes instead $n=1/2$, $k=-1$ and $g_1=3g_0$, \eqref{solffexp}, \eqref{DKmetric} boils
down to the MTZ black hole \cite{Martinez:2004nb}\footnote{In a conformally related frame (the Jordan
frame), the MTZ black hole and various generalizations thereof were discussed in \cite{Nadalini:2007qi}.}.

Another interesting solution with scalar hair was obtained in \cite{Anabalon:2012ta,Feng:2013tza}.
In these papers a general ansatz for the metric and the scalar field are presented, and the expression of the potential is derived by requiring the equations of motion to hold. Although the produced potential in general fails to have a physical motivation, a subclass of it ($\alpha = 0$ in their notation) is inspired by
supergravity. Tailoring to the present normalization (\ref{action}), the four-dimensional solution in \cite{Feng:2013tza} reads 
\begin{align}
\label{FLW}
\D s^2=-\frac{f(r)}{H_1^{1+\mu}(r)H_2^{1-\mu}(r)}\D t^2+
H_1^{1+\mu}(r)H_2^{1-\mu}(r)\left(\frac{\D r^2}{f(r)}+r^2 \D \Sigma_k^2\right)\,,
\end{align}
where 
\begin{align}
\label{Hfphi}
H_i =1+\frac{q_i}r \,, \qquad f = k H_1H_2+g^2 r^2 (H_1^{1+\mu}H_2^{1-\mu})^2 \,, \qquad 
\phi=\frac {\nu}{\sqrt 2}  \ln \left(\frac{H_1}{H_2}\right)\,.
\end{align}
Their scalar potential is given by
\begin{align}
\label{}
V=-\frac 12 g^2 e^{\sqrt 2 (\mu-1)\phi/\nu} \left[
(\mu-1)(2\mu-1)e^{2\sqrt 2 \phi/\nu}-4(\mu^2-1)e^{\sqrt 2\phi/\nu}
+(\mu+1)(2\mu+1)
\right]\,.
\end{align}
Here $\mu, \nu $ are constants satisfying $\mu^2+\nu^2=1 $.
If we set 
\begin{align}
\label{munu}
\mu =\pm (n-1)\,, \qquad \nu =\pm \sqrt{n(2-n)}\,, 
\end{align}
we can recover our potential (\ref{V}) with $g_0=gn/4$, $g_1=g(2-n)/4$ \cite{Lu:2013eoa}. 
The solution (\ref{FLW})  does not have an event horizon for $k\ge 0$, but is very similar to ours.
In fact, taking e.g.~the plus sign in (\ref{munu}),  one sees that \eqref{FLW} is of the form \eqref{DKmetric}
with the $f_i$ proportional to $rH_i$. However, it turns out that the sign of $\phi$  in (\ref{Hfphi}) and (\ref{solffexp}) disagrees. Moreover, the function $\psi$ that results from casting \eqref{FLW} into
the form \eqref{DKmetric} is different from $\psi$ in \eqref{solffexp}. It follows that the static solution
(\ref{FLW}) corresponds to a branch different from ours. It would be very interesting to seek for a
solution that comprises both \eqref{solffexp}, \eqref{DKmetric} and \eqref{FLW}, \eqref{Hfphi}.
An ansatz for $\psi$ and $\phi$ that in principle does this job would be
\begin{align}
\label{generalansatz}
e^{2\psi}=f_1(r)f_2(r)[c_1 r^2+c_2 r+c_3 + c_4 f_1(r)^{2n-1}f_2(r)^{3-2n}]\,, \qquad 
\lambda_n \phi=\ln \left(c_0 \frac{f_2(r)}{f_1(r)}\right)\,,
\end{align}
with $f_i (r)=a_i r+b_i$, where $a_i, b_i, c_i$ are constants. Indeed, our solution has $c_4=0$,
while \eqref{FLW} would correspond to $c_1=c_2=0$. Unfortunately it turns out that \eqref{generalansatz}
works either for \eqref{solffexp}, \eqref{DKmetric} or for \eqref{FLW}, \eqref{Hfphi}, but is unable to
synthesize both. We shall leave the construction of such a more general solution for future work.

Let us finally compare with the spherically symmetric numerical solution discussed in \cite{Hertog:2004bb}.
Our Lagrangian reduces to theirs when $n=1$ with $g_0=g_1$. Hence one may hope that our solution represents their numerical solution.  However, our metric does not admit a spherical horizon when $n=1$, as we will see in the next section. This fact also suggests to look for more general solutions that contain
both \eqref{solffexp}, \eqref{DKmetric} and the numerical one of \cite{Hertog:2004bb}.

\section{Physical discussion}
\label{sec:physicaldiscussion}

In this section, we explore various properties of the hairy black hole 
obtained in the previous section.

\subsection{Mass}
\label{sec:mass}

The mass of a given spacetime is the most fundamental physical quantity. Let us therefore try to identify
a well-defined mass for our spacetime. Since the conserved charges defined at infinity are usually encoded
in the leading terms of physical fields departing from the background spacetime, we consider the
asymptotic behavior of the metric and the scalar field. For this purpose, it is convenient to define the areal
radius by
\begin{equation}
\label{radius}
\rho = \sqrt{f_1^n f_2^{2-n}}\,.
\end{equation}
In terms of $\rho$, the asymptotic expansion ($\rho \to \infty$) of the metric \eqref{DKmetric} reads
\begin{equation}
\D s^2 \simeq -\left(k + \frac{\rho^2}{\ell^2} - \frac{2\mu_1}{\rho}\right)\D \tau^2
+ \left(k + \gamma + \frac{\rho^2}{\ell^2} - \frac{2\mu_2}{\rho}\right)^{-1}
\D\rho^2 + \rho^2\D\Sigma_k^2\,,
\label{asy_metric}
\end{equation}
where $\tau =\rho_0^{-1} t$, 
$\ell=\rho_0/(2\sqrt 2g_0)$ is the AdS radius given in (\ref{rho0}) and
\begin{align}
\label{gamma}
\gamma \equiv \frac{32g_0^2\beta^2}{n(2-n)}\,. 
\end{align} 
We have also defined the two parameters
\begin{subequations}
\begin{align}
\label{mu12}
\mu_1 &=\frac{1}{12}\rho_0\lambda_n^6 (n-1)\beta [3kn^2(n-2)^2+128g_0^2\beta^2(3-2n)(1-2n)] \,,  \\
\mu_2&=\frac{1}{12}\rho_0\lambda_n^6 (n-1)\beta [3kn^2(n-2)^2+128g_0^2\beta^2(5n^2-10n+3)] \,.
\end{align}
\end{subequations} 
The scalar field behaves according to (\ref{scalar_behavior}),  
\begin{align}
\label{phi_ex}
\phi \simeq \phi_1 +\frac{\phi_-}\rho +\frac{\phi_+}{\rho^2 } + {\cal O}(1/\rho^3) \,,
\end{align}
where $\phi_1$ is defined by (\ref{extrema}) and 
\begin{align}
\label{phi_+-}
\phi_-=-2 \lambda_n \beta \rho_0 \,, \qquad 
\phi_+ =-2 (n-1)\lambda_n^3 \beta ^2 \rho_0^2\,. 
\end{align}
It follows that the parameter $\alpha$ in (\ref{scalar_behavior2}) characterizing the boundary condition 
reads
\begin{align}
\label{}
\alpha = \frac 12 (1-n) \lambda_n \,. 
\end{align}
We also note the useful relation
\begin{align}
\label{gamma_rel}
\gamma =\frac 1{2\ell^2} \phi_-^2 \,. 
\end{align}
Thus far, various notions of asymptotically AdS spacetimes have been
defined \cite{Abbott:1981ff,Ashtekar:1984zz,Henneaux:1985tv,Hollands:2005wt}, and lots of apparently
distinct definitions of conserved charges have been proposed. Unfortunately, asymptotic AdS boundary
conditions put forward in these papers do not allow a class of metrics behaving like (\ref{asy_metric}). Due
to the fact that the mass of the scalar field lies in the range (\ref{BFrange}), the stress tensor of the slowly
decaying mode $\phi_-$ of the scalar field does not fall off sufficiently rapidly at infinity, giving rise to a
backreaction to the geometry which modifies the $g^{\rho\rho} $ behaviour from the standard asymptotic
form with $\gamma=0$. This is a main obstruction to construct the conserved quantity following the
prescriptions in \cite{Abbott:1981ff,Ashtekar:1984zz,Henneaux:1985tv,Hollands:2005wt}. This property is
also encoded into the behavior of the Misner-Sharp energy \cite{Misner:1964je}, which is a well-defined
quasi-local energy in pseudo-spherical symmetry \cite{Maeda:2007uu} and is defined by
\begin{align}
\label{}
M_{\rm MS}=\frac 12 \rho \left[k-(\nabla\rho)^2 +\frac{\rho^2}{\ell^2}\right] \,. 
\end{align}
One easily sees that it diverges as $\rho \to \infty$ for our spacetime (\ref{DKmetric}). 

Hertog and Maeda proposed a possible way to avoid this problem \cite{Hertog:2004dr}\footnote{Note that
the result of Hertog and Maeda follows also from holographic renormalization. In particular, mixed
boundary conditions shift the holographic stress tensor according to table 3 in \cite{Papadimitriou:2007sj}.
Computing the conserved charges with this shifted stress tensor yields the charges
of \cite{Hertog:2004dr}. We would like to thank I.~Papadimitriou for pointing out this to us.}.
They generalized
the boundary conditions of Henneaux-Teitelboim \cite{Henneaux:1985tv} in such a way that the relaxed
boundary conditions (i) include spacetimes containing a scalar field with mass parameter in the BF range,
(ii) are invariant under the asymptotic symmetries (eq.~(2.14) of \cite{Hertog:2004dr}), and (iii) allow
charges generated by the corresponding asymptotic symmetries to remain finite. Focusing on the
asymptotically globally AdS case ($k=1$), they obtained \cite{Hertog:2004dr}
\begin{align}
\label{}
 Q[\xi]  =Q_{\rm HT} [\xi] +\frac{\lambda_- }{16\pi \ell} \oint \D \Omega_2 
{\xi^\bot}r^2 \left(\phi^2+\frac{2}3\alpha(\lambda_+-\lambda_-)\phi^{3/\lambda_-}
 \right)\,, 
\end{align}
where $\xi $ is an asymptotic symmetry and $\xi^\bot$ denotes its normal component to the spacelike
surface. $Q_{\rm HT} [\xi] $ denotes the contribution of the Henneaux-Teitelboim
charge \cite{Henneaux:1985tv}. We would like to stress that this piece is divergent, which is precisely
cancelled by the divergence of the second term. In the present case, the mass is given by
$M=Q[\rho_0 \partial_t]$, i.e.\footnote{In refs.~\cite{Lu:2013ura,Liu:2013gja,Lu:2014maa}, it was argued
that an extra work term should contribute to the (variation of the) Hamiltonian defined by Wald's covariant
phase space method \cite{Iyer:1994ys,Wald:1999wa} as $\delta H_\xi=\delta M-X\delta Y$. However, this
additional term $X\delta Y$ violates the necessary condition
$(\delta_1\delta_2-\delta_2 \delta_1) H_\xi=0$ for the existence of a Hamiltonian, unless $X, Y$ are
functionally dependent (in the analysis of \cite{Henneaux:2006hk}, a similar restriction arises from the
consistency of asymptotic symmetry). In the present case, this additional term
vanishes thanks to the relation (\ref{phi_+-}). Notice also that in \cite{Wen:2015uma}, a definition of
mass was proposed
that is similar in spirit to both Wald's formula and the Hamiltonian formalism, with the difference that
\cite{Wen:2015uma} requires the variation of the mass to have no contribution from the variation of
the matter charges.},
\begin{align}
\label{Mass}
M=\mu_2-\frac{32}3 g_0^2\rho_0 n(2-n)(1-n) \lambda_n^6 \beta^3=
\mu_1\,, 
\end{align}
where the last equality was evaluated by using (\ref{mu12}). The constant of integration has been fixed so that it vanishes for AdS spacetime. 
This expression also coincides with the formula given in \cite{Henneaux:2006hk},
\begin{align}
M=\mu_2-\frac 13 m^2 \alpha \phi_- ^{3/\lambda_-}\,, 
\end{align}
where $m$ is the mass (\ref{mass}) at the AdS vacuum $\phi=\phi_1$. 
A natural generalization to the topological cases ($k=0,-1$) is inferred to be 
\begin{align}
\label{}
M=\frac{\Sigma_k \mu_1}{4\pi}\,, 
\end{align}
where $\Sigma_k$ is the area of a unit maximally symmetric space with curvature $k$.  
We will justify this expression below by deriving the first law of black hole thermodynamics. 

The $n=1$ case is particularly interesting since the mass vanishes, whereas the 
scalar field profile remains nontrivial. This does not contradict the positive mass theorem, 
since the $n=1$ spherical solution describes a naked singularity as shown in section \ref{sec:horizon}. 

Note that the parameter $\mu_1$ indeed corresponds to the Coulomb part of the spacetime. 
Namely the Weyl scalar of the Newman-Penrose formalism behaves as
\begin{align}
\label{}
\Psi_2 =-\frac{\mu_1}{\rho^3} + {\cal O}(1/\rho^4)\,.
\end{align}
This behaviour partly justifies the use of the Ashtekhar-Magnon definition of conserved
charges \cite{Ashtekar:1984zz} for spacetimes with scalar field in the BF range, since their mass
essentially corresponds to the coefficient of the ${\cal O}(1/\rho^3)$ term of the electric part of the Weyl tensor. Considering the fall-off behaviour of the stress-energy tensor in \cite{Ashtekar:1984zz}, there
does not appear to be an a priori reason why the Ashtekhar-Magnon charge is finite in the present case.
Perhaps it appears that some miraculous cancellation occurs between terms arising from the metric and the
scalar field.

\subsection{Horizon structure}
\label{sec:horizon}

There appear two possible Killing horizons $r_\pm$,
which are identified by the two roots of $e^{2\psi}=0$, i.e., 
\begin{align}
\label{rpm}
r_\pm = \frac{8g_0\beta(1-n)\pm \sqrt{-2k n^2(2-n)^2-64g_0^2\beta^2 (3-8n+4n^2)}}{4g_0n(2-n )}\,.
\end{align}
$r=r_+$ describes the event horizon of a black hole, on which we focus hereafter. 
The other roots $r=-2\beta/n$ and $r=2\beta/(2-n)$ are not our concern, since they 
correspond to a curvature singularity. This can be easily seen by noting that the areal radius $\rho$
defined by (\ref{radius}) vanishes at these points. 

Let us first discuss the conditions under which spherical horizons exist. 
Setting $k=1$ in \eqref{rpm},
the expression under the square root is nonnegative for $1/2<n<3/2$ and
\begin{equation}
32g_0^2\beta^2 \ge -\frac{n^2(2-n)^2}{3-8n+4n^2}\,. \label{bound_beta}
\end{equation}
In addition, we must impose
that the curvature singularities at $r=-2\beta/n$ and $r=2\beta/(2-n)$ be hidden behind the horizon.
It turns out that this yields a lower bound on $g_0^2\beta^2$ that is more stringent than
\eqref{bound_beta}. A straightforward calculation shows that spherical black holes exist for
\begin{equation}
\frac12 < n < 1 \quad \text{and} \quad g_0\beta > \frac{n(2-n)}{8\sqrt2\sqrt{(2n-1)(1-n)}}
\end{equation}
or
\begin{equation}
1 < n < \frac32 \quad \text{and} \quad g_0\beta < -\frac{n(2-n)}{8\sqrt2\sqrt{(3-2n)(n-1)}}\,.
\end{equation}
Note that for $n=1$ and $k=1$ the solution describes a naked singularity, since in this case either
the dangerous radius $2\beta$ (for $\beta>0$) or $-2\beta$ (for $\beta<0$) is larger than $r_+$.
In the $k=1$ case, the inner horizon $r=r_-$ is always smaller than 
${\rm max}[-2\beta/n,2\beta/(2-n)]$. This means that the singularity is spacelike and the causal structure is the same as for the $k=1$ Schwarzschild-AdS black hole.  

An analogous computation shows that flat horizons ($k=0$) are possible for $1/2<n<1$ and
$\beta>0$ or for $1<n<3/2$ and $\beta<0$. In the special case $n=1$ the horizon always
coincides with one of the singularities. For the planar case, the inner horizon $r_-$ cannot be larger than
the singularity at ${\rm max}[-2\beta/n,2\beta/(2-n)]$.

Finally, hyperbolic black holes ($k=-1$) exist if $\beta$ satisfies the following conditions:
\begin{itemize}
\item $0<n<1/2$:
\begin{equation}
-\frac{n(2-n)}{8\sqrt2\sqrt{(n-1)(2n-3)}} < g_0\beta \le \frac{n(2-n)}{4\sqrt2\sqrt{(2n-1)(2n-3)}}\,,
\label{hyp_horizon_0<n<1/2}
\end{equation}
\item $1/2\le n<1$:
\begin{equation}
g_0\beta > -\frac{n(2-n)}{8\sqrt2\sqrt{(n-1)(2n-3)}}\,,
\end{equation}
\item $1<n\le 3/2$:
\begin{equation}
g_0\beta < \frac{n(2-n)}{8\sqrt2\sqrt{(n-1)(2n-1)}}\,,
\end{equation}
\item $3/2<n<2$:
\begin{equation}
-\frac{n(2-n)}{4\sqrt2\sqrt{(2n-1)(2n-3)}} \le g_0\beta < \frac{n(2-n)}{8\sqrt2\sqrt{(n-1)(2n-1)}}\,.
\label{hyp_horizon_3/2<n<2}
\end{equation}
\end{itemize}
For $n=1$ there is no restriction on $\beta$.
When (\ref{hyp_horizon_0<n<1/2}) or (\ref{hyp_horizon_3/2<n<2}) are satisfied
the inner horizon $r_-$ also appears, 
since it is larger than ${\rm max}[-2\beta/n,2\beta/(2-n)]$.

In the case $k=-1$, $r_+$ and $r_-$ can actually coincide. One finds that extremal hyperbolic hairy
black holes exist when
\begin{equation}
0 < n < \frac12 \quad \text{and} \quad g_0\beta = \frac{n(2-n)}{4\sqrt2\sqrt{(3-2n)(1-2n)}}
\end{equation}
or
\begin{equation}
\frac32 < n < 2 \quad \text{and} \quad g_0\beta = -\frac{n(2-n)}{4\sqrt2\sqrt{(3-2n)(1-2n)}}\,.
\end{equation}
The extremal solution interpolates between AdS$_4$ at infinity and $\text{AdS}_2\times\text{H}^2$
at the horizon. Interestingly, the value of the scalar field at the horizon agrees with the other critical point
of the potential at $\phi=\phi_2$. It follows that the scalar field goes thereby from the supersymmetric
vacuum $\phi_1$ at $r=\infty$ to $\phi=\phi_2$ at the horizon. We explicitly checked that the
near-horizon geometry with $\phi=\phi_2$ breaks supersymmetry. From the holographic point of view,
the black hole describes a ($2+1$)-dimensional superconformal field theory that is perturbed by a
multi-trace deformation and flows to a conformal quantum mechanics in the IR (`flow across
dimensions') \cite{Witten:2001ua,Papadimitriou:2007sj}.

\subsection{Thermodynamics}

Let us next discuss thermodynamic properties of the black holes.
In terms of the horizon radius $r_+$, the area of the black hole horizon reads 
\begin{align}
\label{}
A=\Sigma_k f_1^n(r_+)f_2^{2-n}(r_+) \,.
\end{align}
From (\ref{asy_metric}), the unit time translation at AdS infinity corresponds to 
$\xi=\rho_0 \partial/\partial t$, which is timelike outside and null on the event horizon.
The surface gravity, $\kappa^2=-\frac 12\nabla_\mu\xi_\nu\nabla^\mu\xi^\nu$, is then easily
computed to give 
\begin{align}
\label{}
\kappa = \left.\frac{(e^{2\psi})'}{2\sqrt 2 g_0^{(-2+n)/2}g_1^{(2-n)/2}n^{-n/2}(2-n)^{(n-2)/2}f_1^nf_2^{2-n}}\right|_{r=r_+} \,,
\end{align}
which accords with the Euclidean prescription of removing conical singularities. 
The Hawking temperature is then defined by $T=\kappa/2\pi$. 
An elementary computation reveals that the first law of black hole
thermodynamics \cite{Bardeen:1973gs} holds,
\begin{align}
\label{}
\delta M= \frac{\kappa}{8\pi}\delta A \,. 
\end{align}
It turns out that the scalar charge fails to contribute to the 1st law, consistent with the analysis 
in  \cite{Papadimitriou:2005ii}. Due to the appearance of a curvature scale, there is no straightforward 
expression for an integrated mass formula. 

\begin{figure}[t]
\begin{center}
\includegraphics[width=15cm]{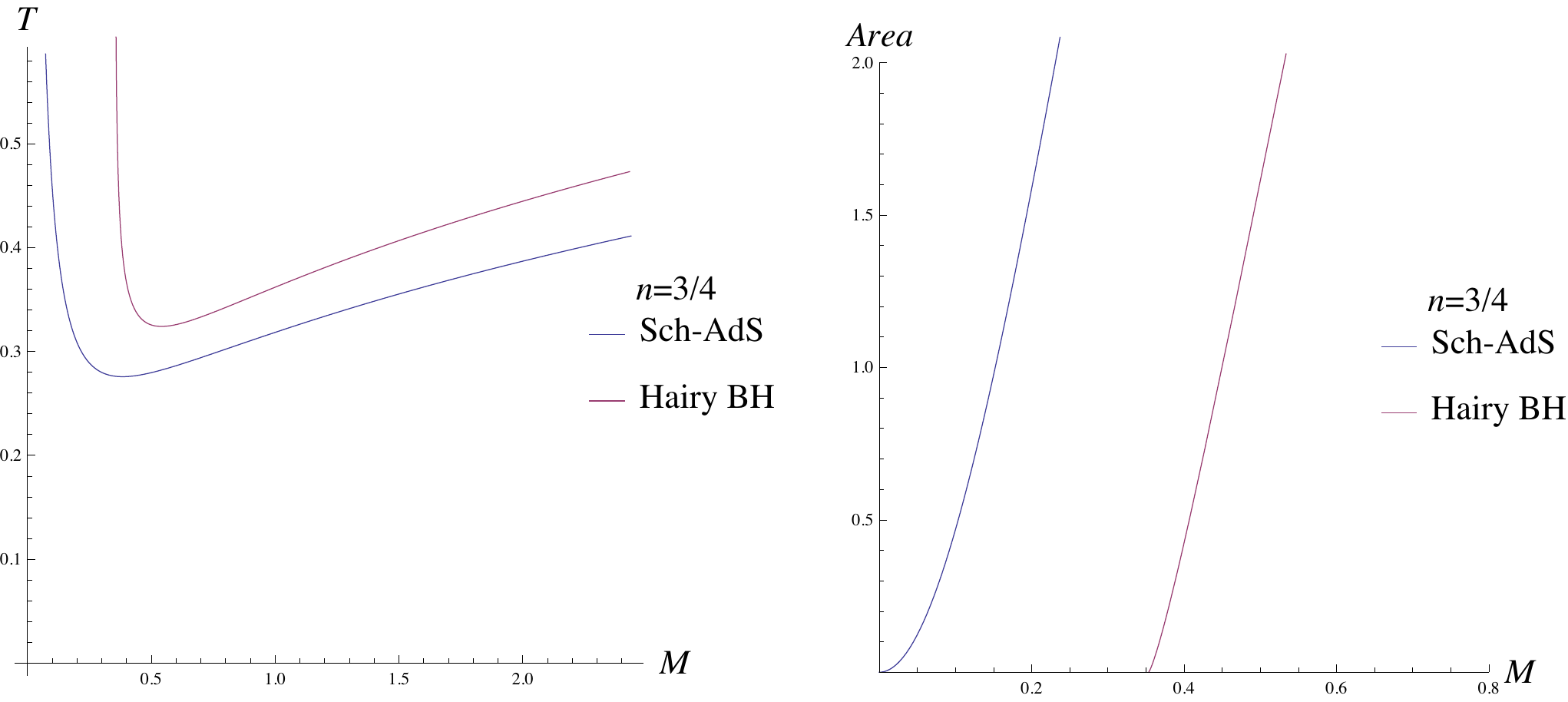}
\caption{The behaviour of temperature and area as functions of the mass for $k=1$, 
$n=3/4$. The asymptotic AdS radius has been set to $\ell=1$ and the corresponding  
quantities for the Schwarzschild-AdS black hole are also shown. 
Both black holes have a critical temperature. 
Our black hole solution has a positive mass bound and a smaller entropy 
compared to the Schwarzschild-AdS black hole. 
}
\label{fig:entropy}
\end{center}
\end{figure}

Fig.~\ref{fig:entropy} shows the temperature and the horizon area as functions of the mass for $k=1$
and $n=3/4$ (in red), as compared to the Schwarzschild-AdS solution (blue).
One can see that the qualitative behaviour is very similar to that of the Schwarzschild-AdS black hole. 
There appears a critical temperature $T_{\text c}$ above which we have two kinds of black holes
with $M\lessgtr M_{\text c}$ \cite{Hawking:1982dh}. 
We do not write down the explicit expression for $T_{\text c}$ since it is quite messy, but it can be easily
computed. The behaviour of the specific heat $C=T\partial S/\partial T$ is also similar to the
Schwarzschild-AdS black hole, since one has $C\lessgtr 0$ for $M\lessgtr M_{\text c}$ and
$C\to\pm\infty$ for $M\to M_{\text c}\pm 0$.

One sees from fig.~\ref{fig:entropy} that, for a given mass, the area of the hairy black hole is always
smaller than the one of the Schwarzschild-AdS black hole, which implies that the hairy black hole is
unstable. Such an instability was actually also found for the numerical solutions in the
literature \cite{Torii:2001pg,Hertog:2004bb}. We shall verify in section \ref{sec:instability} that our
solution admits also an unstable mode.

In the case of a hyperbolic horizon ($k=-1$), there appears an inner horizon
outside the curvature singularities. 
The area product of the two Killing horizons is given by 
\begin{align}
\label{areaproduct}
A_+A_-=&\frac{\Sigma_k^2}{256g_0^4n^4}\left(\frac{n}{2-n}\right)^{2n}
\left[kn^2(2-n)^2+128 g_0^2(1-n)(3-2n)\beta^2\right]^n\nonumber \\
&\times \left[kn^2(2-n)^2+128 g_0^2(1-n)(1-2n)\beta^2\right]^{2-n} \,. 
\end{align}
According to the analysis in \cite{Cvetic:2010mn}, the area product of a generic black hole 
depends only on the (quantized) charges, the angular momenta and the cosmological constant. 
However, the right hand side of (\ref{areaproduct}) does not correspond to such a
quantity\footnote{Visser argued that the contribution coming from the virtual horizons should be taken
into account \cite{Visser:2012wu}. However, such contributions do not make sense in the present context,
since other two roots of $e^{2\psi}=0$ correspond to the singularities at which the area vanishes.}. 
It would be interesting to explore the physical meaning of the right hand side of (\ref{areaproduct}).

\subsection{Instability against radial perturbations}
\label{sec:instability}

The behaviour of the entropy of our hairy black hole with respect to Schwarzschild-AdS
(fig.~\ref{fig:entropy}) implies that our solution is unstable against dynamical perturbations. Focusing on
spherically symmetric perturbations, we show that this is indeed the case. 

The background spacetime we consider is the spherical solution ($k=1$) of the form
\begin{align}
\label{}
g_{\mu\nu}^{(0)}\D x^\mu \D x^\nu =-f(r)\D t^2+\frac{\D r^2}{f(r)} +\rho(r)^2 (\D \theta^2+\sin^2\theta \D \varphi^2) \,, \qquad \phi^{(0)}=\phi^{(0)}(r) \,,
\end{align}
where 
$f=e^{2\psi}/(f_1^n f_2^{2-n})$ and $\rho (r)$ is given by (\ref{radius}). 
Here and in what follows, we attach $(0)$ to denote the background quantities. 
The governing equations are 
\begin{align}
\label{}
E^\mu{}_\nu\equiv R^\mu{}_\nu-\frac 12 R \delta ^\mu{}_\nu -T^\mu{}_\nu =0\,, \qquad E_\phi \equiv 
\nabla^\mu\nabla_\mu\phi-\partial_\phi V=0 \,.
\end{align}
Let us consider the spherically symmetric perturbations
\begin{align}
\label{}
  g_{\mu\nu}\simeq  g^{(0)}_{\mu\nu}+g^{(1)}_{\mu\nu}(r) e^{-i\omega t}  \,, \qquad 
  \phi\simeq  \phi ^{(0)}+\phi^{(1)} (r)  e^{-i\omega t} \,, 
\end{align}
where $(1)$ corresponds to the perturbed value.  
Using the gauge freedom 
$g^{(1)}_{\mu\nu}\to g^{(1)}_{\mu\nu}+\mas L_\xi g_{\mu\nu}^{(0)}$, 
we can work in a gauge where only $g^{(1)}_{tt}$ and $g^{(1)}_{rr}$ are nonvanishing
(see \cite{Ishibashi:2011ws} for a review of gravitational perturbations in spherically symmetric spacetimes). 

Using the background Einstein's equations $E^{(0)}_{\mu\nu}=0$ 
and linearized Einstein's equations $E^{(1)}_{\mu\nu}=0$, 
the perturbed scalar equation $E^{(1)}_\phi=0$ reduces to the single master equation
\begin{align}
\label{Schrodinger_eq}
\left(-\frac{\D ^2}{\D r_*^2}+U(r)\right)\Phi =\omega ^2 \Phi \,, 
\end{align}
where $\Phi=\rho \phi^{(1)}$ and $r_* =\int \D r/f(r)$ denotes the tortoise coordinate.
The potential $U(r)$ reads
\begin{align}
\label{}
U=\frac{f}{2\rho \rho'{}^2}[f\rho^3(\phi^{(0)})'{}^4+2 \rho'^2(f\rho')'
-2\rho \rho'(f\rho)'(\phi^{(0)})'{}^2
+4\rho^2 \rho'(\phi^{(0)})'V_\phi+2\rho \rho'{}^2 V_{\phi\phi}] \,,
\end{align}
where the prime denotes the differentiation with respect to $r$, 
$V_\phi=\partial_\phi V$ and $V_{\phi\phi}=\partial_\phi^2 V$ correspond to the background value.
The tortoise coordinate asymptotically behaves as 
\begin{align}
\label{}
r_* \sim \frac {1}{f'(r_+)} \ln (r-r_+) \to -\infty ~~(r\to r_+ ) \,, 
\qquad 
r_*\sim -\frac {\ell^2}r + K~~(r\to \infty)\,, 
\end{align}
where $K$ is a constant. 

\begin{figure}[t]
\begin{center}
\includegraphics[width=8cm]{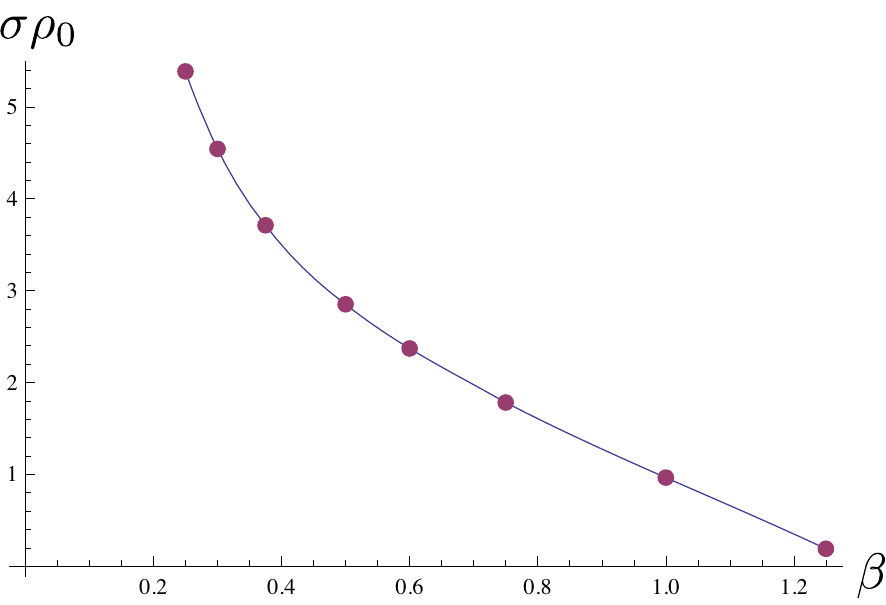}
\caption{Instability rate against radial perturbations for $n=3/4$, $g_0 =\ell=1$.}
\label{fig:instability}
\end{center}
\end{figure}

We are now looking for an unstable mode which occurs at a purely imaginary frequency.
Hence we set $\sigma =-i\omega $ ($\sigma >0$) as in ref.~\cite{Hertog:2004bb}. 
Since $U \to 0$ for $r_*\to -\infty$, the normalizable asymptotic solution at the horizon reads
\begin{align}
\label{BC_horizon}
\Phi \sim \exp (\sigma r_* ) \,, \qquad (r_* \to -\infty)\,. 
\end{align}
Let us next consider the boundary condition for $\Phi$ at infinity. 
The scalar field behaves as 
\begin{align}
\label{}
\phi \sim \frac{\phi_-}{\rho} +\frac{\alpha \phi_-^2}{\rho^2 }\,.
\end{align}
Hence for $\phi =\phi^{(0)} +\phi^{(1)}e^{-i\omega t}$ we have 
\begin{align}
\label{phi1_bc}
\phi^{(1)} \sim  \frac{\phi_-^{(1)}}{\rho} +\frac{2\alpha \phi_-^{(0)}\phi_-^{(1)}}{\rho^2 }\,.
\end{align}
Let $U_0=U (r\to \infty)<0$ and suppose 
$\ti \sigma =\sqrt{-U_0-\sigma^2}$ is positive. 
Then the asymptotic solution reads
\begin{align}
\label{Phi_asysol_inf}
\Phi \sim A \cos(\ti\sigma r_*+b)\simeq A \left[
\cos (\ti \sigma K+b)+\frac{\ti \sigma \ell^2}r \sin (\ti \sigma K+b)
\right] \,, 
\end{align}
where $A$ and $b$ are constants. 
Noting $\rho \sim \rho_0 (r+\ma O(1/r))$, a comparison of (\ref{phi1_bc}) and (\ref{Phi_asysol_inf}) gives 
\begin{align}
\label{}
\ti \sigma  \ell^2 \tan (\ti \sigma K+b) =\frac{2\alpha \phi_-^{(0)}}{\rho_0} \,. 
\end{align}
This translates into
\begin{align}
\label{BC_infinity}
\frac{\D }{\D r_*} \Phi\simeq -\frac{2\alpha \phi_-^{(0)}}{\rho_0\ell^2}\Phi  \quad (r\to \infty)\,. 
\end{align}
We now have a Schr\"odinger-type equation (\ref{Schrodinger_eq}) 
with the boundary conditions (\ref{BC_horizon}), (\ref{BC_infinity}). 
We numerically solved the eigenvalue problem and found an unstable mode as displayed  in 
fig.~\ref{fig:instability}. As $\beta $ increases, the instability rate becomes smaller. 
Since we fix the AdS curvature, large $\beta $ means large horizon radius. 
Thus the smaller hairy black hole is more unstable. For the chosen values
$n=3/4$, $g_0=\ell =1$, the unstable mode seems to disappear around $\beta \sim 1.4$. 
The profound physical reason for this value and its relation to thermodynamics remain still to be understood.

\subsection{Asymptotically de Sitter case}

A Wick rotation of the coupling constants,
$g_0\to i g_0$, $g_1\to i g_0$, reverses the sign of the scalar potential, 
hence the metric is asymptotically de~Sitter. The potential is no longer expressed in terms of 
a real superpotential, yet the solution still solves the equations of motion (\ref{eom}). 
In eqs.~(\ref{solffexp}), the expressions of 
$f_1$, $f_2$ and $\phi$ are left invariant, whereas 
\begin{align}
\label{}
e^{2\psi}=&\left(r+\frac{2\beta}{n}\right)\left(r-\frac{2\beta}{2-n}\right)\nonumber \\ & \times\left[k-8 g_0^2 \left(r^2-\frac{4(1-n)}{n(2-n)}\beta r+4\frac{5n^2-10n+4}
{n^2(2-n)^2}\beta^2 \right)\right] \,. 
\end{align}
Our solution is compatible with the no-hair theorem for asymptotically de Sitter spacetimes~\cite{Bhattacharya:2007zzb} since the potential is not convex. 

The Killing horizons are given by
\begin{subequations}
\begin{align}
\label{rpc}
r_{\text c} = &\frac{8g_0\beta(1-n)+\sqrt{2k n^2(2-n)^2-64g_0^2\beta^2 (3-8n+4n^2)}}{4g_0n(2-n )}\,,\\
r_+ = &\frac{8g_0\beta(1-n)-\sqrt{2k n^2(2-n)^2-64g_0^2\beta^2 (3-8n+4n^2)}}{4g_0n(2-n )}\,.
\end{align}
\end{subequations}
$r=r_{\text c}$ is a cosmological horizon, while $r=r_+\le r_{\text c}$ is a black hole event horizon. 
It turns out that the condition for the existence of these real roots is weaker than the no naked singularity 
condition $r_+ \ge {\rm max}[2\beta/(2-n), -2\beta/n]$. One finds that horizons exist only for $k=1$ 
with the following parameter region:
\begin{itemize}
  \item $0<n\le 1/2$:
  \begin{align}
\label{}
\frac{n(2-n)}{8\sqrt{2(1-n)(1-2n)}}< g_0\beta \le 
\frac{n(2-n)}{4 \sqrt{2(1-2n)(3-2n)}}\,, 
\end{align}
  \item $3/2\le n<2$:
    \begin{align}
\label{}
-\frac{n(2-n)}{4 \sqrt{2(2n-3)(2n-1)}}\le g_0\beta <
-\frac{n(2-n)}{8\sqrt{2(n-1)(2n-3)}}\,.
\end{align}
\end{itemize}
For these parameter ranges, the global structure is the same as for the $k=1$ Schwarzschild-de~Sitter
black hole. One can also verify that there exist no lukewarm black holes for which the temperatures of
the black hole horizon and the cosmological horizon are equal.

\section{Final remarks}
\label{final-rem}

In this paper we constructed a new family of black holes with scalar hair in $N=2$ Fayet-Iliopoulos gauged supergravity. A distinguished feature of our solution is that it is sourced only with a scalar field. This kind of exact black hole solution  is of importance and will be a stepping stone for constraining the conditions under which the no-hair theorem is valid. 

We explored various properties of these black holes. In particular, we pointed out that the standard
methods of computing conserved charges does not work, since the asymptotic behaviour of the metric is
out of the framework of `asymptotically AdS' proposed in the literature. Nevertheless, we identified a
well-defined mass function following the prescription of Hertog and Maeda~\cite{Hertog:2004dr}. This fixes some confusions prevailing in the literature. 

One may be tempted to hope that this solution would be a novel counterexample to the 
no hair conjecture. Since our black hole is unstable, this is not the case. However, this instability is
interesting from the holographic viewpoint \cite{Gubser:2000mm}. It would be nice to see if our solution
has some potential applications in the AdS/CFT correspondence, and in particular to condensed matter
physics, where one typically includes the leading scalar operator in the dynamics. This is generically
uncharged, and is dual to a neutral scalar field in the bulk.

Another possible extension of the present work is to look for rotating black holes. In this case, there
might exist a rotating black hole which admits only a helical Killing vector \cite{Herdeiro:2014goa}. It
would be interesting to construct this kind of exact solutions in the framework of supergravity.

\section*{Acknowledgements}

This research was supported in part by INFN and JSPS.

\end{document}